\documentstyle[twoside,fleqn,espcrc2]{article}

\newcommand{\AmS}{{\protect\the\textfont2
  A\kern-.1667em\lower.5ex\hbox{M}\kern-.125emS}}
  
\def\be{\begin{equation}}
\def\ee{\end{equation}}
\newcommand{\beqn}{\begin{eqnarray}}
\newcommand{\eeqn}{\end{eqnarray}}
\newcommand{\nn}{\nonumber}
\newcommand{\MSbar}{\overline{\mbox{MS}}}
\newcommand{\bea}{\begin{eqnarray}}
\newcommand{\eea}{\end{eqnarray}}
\title{$B$-parameters for $\Delta S = 2$ Supersymmetric Operators}

\author{
C.R.~Allton\address{Dep. of Physics, University of Wales Swansea, Swansea, UK.}
$\!\!\!,$
L.~Conti\address{Dip. di Fisica, Universit\`a di Roma ``Tor Vergata '' and 
INFN Sezione di Roma II, I-00133 Roma, Italy.}$\hspace{-0.05mm}$\thanks{Talk 
presented by L.~Conti.}
$\!\!\!,$
A.~Donini\address{Dep. de Fisica Teorica, Universidad Autonoma de Madrid, Cantoblanco, E-28049 Madrid, Spain.}
$\!\!\!,$
V.~Gimenez\address{Dep. de Fisica Teorica and IFIC, Universidad de Valencia, Burjassot, E-46100 Valencia, Spain.}
$\!\!\!,$
L.~Giusti\address{Scuola Normale Superiore, and INFN, Sezione di Pisa, 56100 Pisa, Italy.}
$\!\!\!,$
G.~Martinelli\address{Dip. di Fisica, Universit\`a di Roma ``La Sapienza'' and INFN, Sezione di Roma I, I-00185 Roma, Italy.}
$\!\!\!,$
M.~Talevi\address{Dep. of Physics \& Astronomy, University of Edinburgh, Edinburgh EH9 3JZ, UK.} 
\mbox{and A.~Vladikas$^b$}}

\begin{document}

\begin{abstract}
We present the first lattice measurement, using Non Perturbative Renormalization Method, of the
B-parameters of the dimension-six four-fermion operators relevant for the 
supersymmetric corrections to the $\Delta S=2$ transitions.
\end{abstract}

\maketitle

\section{INTRODUCTION}
\label{sec:intro}
This work is the first lattice calculation
of the matrix elements of the most 
general set of $\Delta S=2$ dimension-six four-fermion operators, 
renormalized non-perturbatively (NP) in the RI (MOM) 
scheme~\cite{NP}--\cite{4ferm_teo}. 
The main parameters and the details of the simulations are given in
ref.~\cite{bsusy}.
Our results can be  combined with the recent two-loop  calculation of  the anomalous 
dimension matrix in the same renormalization scheme~\cite{scimemi}  to 
obtain $K^{0}$--$\bar K^{0}$ mixing amplitudes which are consistently 
computed at the next-to-leading  order.  A phenomenological application of the results 
for the matrix elements  given below, combined with a complete 
next-to-leading order  (NLO) 
evolution of the Wilson coefficients, can be found in~\cite{inprep}. 

\par 
The $B$-parameter of the matrix element $\langle \bar K^0 \vert O^{\Delta S = 2}
 \vert K^0 \rangle$,  commonly known as $B_K$, has been extensively studied 
 on the lattice; for the other 
operators, instead,   all the phenomenological analyses beyond the 
SM have used 
$B$-parameters equal to one, which in some cases is a very crude approximation.  
Moreover, with respect to other calculations, the systematic errors 
in our results are reduced by using the tree-level improved Clover
action~\cite{sw,heat} and by  renormalizing NP the lattice operators.

\par 
We have used the supersymmetric basis~\footnote{This basis is also the one
for which the  numerical values of the Wilson  coefficients, 
computed at the NLO, have been given in~\cite{inprep}.}:
\bea O_1 &=& \bar s^\alpha \gamma_\mu (1- \gamma_{5} ) d^\alpha \ 
\bar s^\beta \gamma_\mu (1- \gamma_{5} )  d^\beta ,  \nn \\ 
O_2 &=& \bar s^\alpha (1- \gamma_{5} ) d^\alpha \ 
 \bar s^\beta  (1- \gamma_{5} )  d^\beta ,  \nn \\ 
O_3 &=& \bar s^\alpha  (1- \gamma_{5} )  d^\beta  \ 
 \bar s^\beta   (1- \gamma_{5} ) d^\alpha ,  \label{eq:ods2} \\ 
O_4 &=& \bar s^\alpha  (1- \gamma_{5} ) d^\alpha \  
\bar s^\beta  (1 + \gamma_{5} )  d^\beta ,  \nn \\ 
O_5 &=& \bar s^\alpha  (1- \gamma_{5} )  d^\beta \ 
 \bar s^\beta (1 +  \gamma_{5} ) d^\alpha , \nonumber
\eea 
where $\alpha$ and $\beta$ are colour indices. 
The $B$-parameters for these operators are defined as
\bea \langle  \bar K^{0} \vert \hat  O_{1} (\mu) \vert K^{0} 
\rangle \hspace{-1mm}&\hspace{-1mm}=\hspace{-2mm}&\hspace{-2mm}
\frac{8}{3} M_{K}^{2} f_{K}^{2} B_{1}(\mu) , \nn \\
\langle  \bar K^{0} \vert \hat O_{2} (\mu) \vert K^{0} \rangle \hspace{-1mm}&\hspace{-1mm}=\hspace{-2mm}&\hspace{-2mm}
%-\frac{5}{3} \left( \frac{ M_{K} }{ m_{s}(\mu) + m_d(\mu) }\right)^{2}
%M_{K}^{2} f_{K}^{2} B_{2}(\mu) , \nn \\
-\frac{5}{3} \frac{ M_{K}^4 f_{K}^{2} }
                  { \left( m_{s}(\mu) + m_d(\mu) \right)^{2} } B_{2}(\mu) , \nn \\
\langle  \bar K^{0} \vert \hat O_{3} (\mu) \vert K^{0} \rangle \hspace{-1mm}&\hspace{-1mm}=\hspace{-2mm}&\hspace{-2mm}
%\frac{1}{3} \left( \frac{ M_{K} }{ m_{s}(\mu) + m_d(\mu) }\right)^{2}
%M_{K}^{2} f_{K}^{2} B_{3}(\mu) , \label{eq:bpars} \nn \\
\frac{1}{3} \frac{ M_{K}^4 f_{K}^{2} }
                 { \left( m_{s}(\mu) + m_d(\mu) \right)^{2} } B_{3}(\mu) , 
\label{eq:bpars}  \\
\langle  \bar K^{0} \vert \hat O_{4} (\mu) \vert K^{0} \rangle \hspace{-1mm}&\hspace{-1mm}=\hspace{-1mm}&\hspace{-2mm}
%2 \left( \frac{ M_{K} }{ m_{s}(\mu) + m_d(\mu) }\right)^{2}
%M_{K}^{2} f_{K}^{2} B_{4}(\mu) ,\nn \\
2 \frac{ M_{K}^4 f_{K}^{2} } 
       { \left( m_{s}(\mu) + m_d(\mu) \right)^{2} } B_{4}(\mu) ,\nn \\
\langle  \bar K^{0} \vert \hat  O_{5} (\mu) \vert K^{0} \rangle \hspace{-1mm}&\hspace{-1mm}=\hspace{-2mm}&\hspace{-2mm}
%\frac{2}{3} \left( \frac{ M_{K} }{ m_{s}(\mu) + m_d(\mu) }\right)^{2}
%M_{K}^{2} f_{K}^{2} B_{5}(\mu) , \nn  \eea
\frac{2}{3} \frac{ M_{K}^4 f_{K}^{2} }
                 { \left( m_{s}(\mu) + m_d(\mu) \right)^{2} } B_{5}(\mu) , \nn  \eea
where the operators $\hat O_{i}(\mu)$ and  the quark masses   are 
renormalized at the scale 
$\mu$  in the same scheme.  The numerical 
results for the $B$-parameters, $B_{i}(\mu)$, computed in this paper refer
to the RI scheme. 
\section{NON-PERTURBATIVE RENORMALIZATION}
\label{sec:npm}
Because of the breaking of the chiral symmetry with the Wilson
fermion, each operator in the $\Delta S = 2$ Hamiltonian 
mixes with operators belonging to different chiral representations
\cite{MARTIW} so that the correct chiral behaviour is
achieved only in the continuum limit. 
This represented a long-standing problem in the evaluation of 
$B_K$ only recently
solved with the introduction of Non-Perturbative Renormalization methods.
In these approaches the  renormalization constants (mixing matrix) are computed
non-perturbatively on the lattice 
either  by projecting on external quark and gluon
states (NPM) as proposed in ref. \cite{NP} or 
by using chiral Ward Identities \cite{WI,JAPBK}.

\par
Recent studies of the $B_K$ and of the B-parameters of the electro-penguin 
operators, $B_7^{3/2}$ and $B_8^{3/2}$ (which coincide 
with $B_{4}$ and $B_{5}$ respectively),
with both non-perturbative renormalization methods, 
\cite{DS=2}--\cite{contil} and \cite{JAPBK}, 
found that the NP renormalization of the lattice operators
gives $B$-parameters that significantly differ from those renormalized
perturbatively (PT)~\cite{gupta_bp}.
The discretization effects are less important than those due to the PT evaluation
of the mixing coefficients. 

\par
In this work we use the NPM renormalization.
The results for all the renormalization constants for 
the complete basis of four-fermion operators
(computed with the NPM, for  several renormalization scales $\mu$,  at $\beta = 
6.0$ and $6.2$) can be found in \cite{4ferm_teo}.
\section{$B$-PARAMETERS}
\label{sec:res}
\par  The $B$-parameters are usually defined as
\be
B_i(\mu) = 
\frac{ \langle \bar K^0 \vert \hat O_i (\mu) \vert
K^0 \rangle } {\langle \bar K^0 \vert \hat O_i \vert
K^0 \rangle_{VSA} } \,\, ,
\label{eq:bkdef}
\ee
where VSA means Vacuum Saturation Approximation.
The VSA values of the matrix elements of $\hat O_{4}$ and $\hat O_{5}$
differ from the factors appearing in the 
definition of the $B$-parameters in eq.~(\ref{eq:bpars}) by the terms
proportional to $1/3 M^2_K f_K^2$ and $M^2_K f_K^2$ respectively. These terms, which  originate 
from the squared matrix elements of the axial current, are of higher order in the chiral 
expansion and  have been dropped in our definition of the $B$-parameters
eq.~(\ref{eq:bpars}).
This implies that, out of the chiral limit,  the values of 
$B_{4}$ and $B_{5}$  with our definition  differ from those obtained  
by using (\ref{eq:bkdef}). 
Out of the chiral limit, with  the standard definition of the $B$-parameters  obtained by using 
the VSA normalization, the scaling properties of 
$B_{4}(\mu)$  and $B_{5}(\mu)$  would have been much more complicated.
The reason is that, 
in these cases, the VSA has a 
piece which scales as the squared pseudoscalar density and another one
(proportional to the physical quantity $\vert \langle \bar  K^0 \vert \hat 
A_\mu \vert 0 \rangle \vert^2$) which is renormalization group 
invariant. 
The $\mu$-independence of the final result would then have been 
recovered in a very intricate way.
Since the definition of the $B$-parameters is conventional, we prefer 
to use that of eq.~(\ref{eq:bpars}),  for which the scaling properties of all the 
$B$-parameters are the simplest ones.  Moreover, with this choice,
they are the same  as   those  derived in the  chiral limit.

\par
We stress that $B_1 = B^{\Delta S = 2}, B_4 = B^{3/2}_8$ and $B_5 = B^{3/2}_7$. 
In ref.~\cite{contil},  
the results referred to the operators  $O^{\Delta S = 2}, O^{3/2}_8$
and $O^{3/2}_7$ at  $\beta=6.0$ only. In this paper, we present the results
for all the $B$-parameters and for $\beta=6.0$ and $6.2$.
\section{NUMERICAL RESULTS}
\label{sec:numres}
%_____________________________________________________________________
\begin{table*}[thbp]
%\catcode`?=\active \def?{\kern\digitwidth}
\begin{tabular*}{\textwidth}{@{}l@{\extracolsep{\fill}}ccccccccccc}
\hline
& &\multicolumn{1}{c}{ ref.\protect\cite{contil}} & $\hspace{1mm}$ & \multicolumn{7}{c}{ this work}   \\
\cline{3-3}
\cline{5-11}
& &\multicolumn{1}{c}{$\beta=6.0$}& &
\multicolumn{2}{c}{$\beta=6.0$}&$\hspace{1mm}$ & \multicolumn{2}{c}{$\beta=6.2$}&$\hspace{1mm}$ & best estimate  \\
\cline{3-3}
\cline{5-6}
\cline{8-9}
\cline{11-11}
& & $m_{K}=0$ & & $m_{K}=0$ & $m_{K}=m_{K}^{exp}$ & & $m_{K}=0$ & $m_{K}=m_{K}^{exp}$ & & $m_{K}=m_{K}^{exp}$ 
\\
\hline
$B_1 = B_K$       &   & 0.66(11)& & 0.70(15) & 0.70(15)& & 0.68(21)& 0.68(21)&  & 0.69(21)  \\
$B_2$             &   & ---     & & 0.61(3)  & 0.66(3)&  & 0.63(6) & 0.66(4) &  & 0.66(4)   \\
$B_3$             &   & ---     & & 1.10(8)  & 1.12(7)&  & 0.94(16)& 0.98(12)&  & 1.05(12)  \\
$B_4 = B^{3/2}_8$ &   & 1.03(3) & & 1.04(4)  & 1.05(3)&  & 0.98 (8)& 1.01(6) &  & 1.03(6)   \\
$B_5 = B^{3/2}_7$ &   & 0.72(5) & & 0.68(7)  & 0.79(6)&  & 0.46(13)& 0.67(10)&  & 0.73(10)  \\
\hline
\end{tabular*}
\caption{\it{$B$-parameters at the renormalization scale 
$\mu = a^{-1} \simeq 2$~GeV, corresponding to $\mu^{2} a^{2}=0.96$ and $\mu^{2} 
a^{2}=0.62$  at  $\beta=6.0$ and $6.2$ respectively. All results are in 
the RI (MOM) scheme.}}
\label{tab:summary}
\end{table*}
% -----------------------------------------------------
Our simulations have  been performed at $\beta = 6.0$ and $6.2$ with the tree-level Clover
action, for several values of the quark masses,
in the quenched approximation. 
In table \ref{tab:summary}, we summarize our results.

\par   
In constructing the renormalized operators  we have used the central 
values of the renormalization constants neglecting their 
 statistical errors.  

\par   
At $\beta=6.0$, the results for $B_{1}$, $B_{4}$ and $B_{5}$ 
extrapolated to the chiral limit  are slightly  different from those of 
 ref.~\cite{contil}.
There are several  reasons for the differences: 
i) we fix the scale and the strange quark mass using
the lattice-plane method of ref.~\cite{giusti}; 
ii) in the present analysis, we use the ``lattice dispersion 
relation'' instead than the continuum one used in \cite{contil}; 
iii) in 
order to reduce the systematic effects due to  higher order terms  in the 
chiral expansion, i.e. to higher powers of  $p \cdot q$, we 
have not used the results corresponding to  $\vec p=2 \pi /L(1,0,0)$  and 
$\vec q=2 \pi/L(-1,0,0)$.  This choice stabilizes the results for $B_{1}$   
between $\beta=6.0$ and $\beta=6.2$  whilst the results for the 
other $B$-parameters remain essentially unchanged. 
\section{CONCLUSIONS}
\label{sec:concl}
Although  we have data at two different values of the lattice spacing, the 
statistical errors, and the uncertainties in the extraction of the 
matrix elements,  are too large to enable any extrapolation to the continuum limit 
$a \to 0$ : within the precision of our results  we cannot detect the dependence of
$B$-parameters on $a$. For this reason, we estimate the central values
by  averaging the $B$-parameters obtained with the physical mass
$m_K^{exp}$ at the two values of  $\beta$.
Our best estimates are reported in the last column of the table \ref{tab:summary}.
We observe that the  lattice  values  of  $B_{3,4}$ are close to their VSA 
whereas  this is not true for $B_{1,2,5}$.

\par
In ref.~\cite{gupta_bp} $B_2$ and $B_3$ have been obtained at $\beta = 6.0$
with the Wilson action and the operators renormalized perturbatively in the
$\MSbar$ scheme; the result is
$B_2 = 0.59(1) \mbox{ and } B_3 = 0.79(1)$.
Although a direct comparison is not possible (our results are in the RI
scheme), to the extent that the matching coefficients between the two schemes
are a small effect \cite{contil}, comparison of the NP results and the PT ones
suggests that PT renormalization behaves poorly in
some cases. This confirms the need for NP renormalization.

\par
Our  results  allow an improvement in  the accuracy  of phenomenological analyses 
intended to put bounds on basic parameters of theories
beyond the Standard Model.


\begin{thebibliography}{999}
%
\def\edlat{Lattice 97, 15th Int. Symp. on Lattice Field Theory,
Edinburgh, Scotland, 1997}
\def\stlolat{Lattice 96, 14th Int. Symp. on Lattice Field Theory, St
Louis, USA, 1996}
\def\ozlat{Lattice 95, 13th Int. Symp. on Lattice Field Theory,
Melbourne, Australia, 1995}
\def\biellat{Lattice 94, 12th Int. Symp. on Lattice Field Theory,
Bielefeld, Germany, 1994}
\def\txlat{Lattice 93, 11th Int. Symp. on Lattice Field Theory,
Dallas, Texas, 1993}
\def\nllat{Lattice 92, 10th Int. Symp. on Lattice Field Theory,
Amsterdam, Netherlands, 1992}
\def\warsaw{ICHEP96, 28th Int. Conf. on High Energy Physics, Warsaw,
Poland, 25--31 July 1996, edited by Z. Ajduk and A.K. Wroblewski,
World Scientific, Singapore (1997)}
%
\def\prd#1{Phys. Rev. D {\bf #1}}
\def\prl#1{Phys. Rev. Lett. {\bf #1}}
\def\plb#1{Phys. Lett. B {\bf #1}}
\def\npb#1{Nucl. Phys. B {\bf #1}}
\def\npbps#1{Nucl. Phys. B (Proc. Suppl.) {\bf #1}}
\def\npaps#1{Nucl. Phys. A (Proc. Suppl.) {\bf #1}}
\def\zpc#1{Z. Phys C {\bf #1}}
\def\nima#1{Nucl. Instrum. Meth. A {\bf #1}}
\def\cmp#1{Commun. Math. Phys. {\bf #1}}
\def\physrep#1{Phys. Rep. {\bf #1}}
%
\bibitem{NP}
G.Martinelli {\em et al.},Nucl.Phys.{\bf B445}(1995) 81.  
\bibitem{DS=2}
A.~Donini {\em et al.}, Phys.~Lett. {\bf B360} (1996) 83.
\bibitem{B_K}
M.Crisafulli {\em et al.},Phys.Lett.{\bf B369}(1996) 325.
\bibitem{contil} L. Conti {\em et al.}, Phys.~Lett. {\bf B421} (1998) 273.
\bibitem{4ferm_teo}
A.~Donini {\em et al.}, prep. ROME1-1181/97 in preparation.
\bibitem{bsusy}C.R.~Allton {\em et al.}, hep-lat/9806016.
\bibitem{scimemi} M. Ciuchini {\em et al.}, hep-ph/9711402.
\bibitem{inprep} M. Ciuchini {\em et al.}, hep-ph/9808328.
\bibitem{sw} B.~Sheikholeslami and R.~Wohlert, \npb{259} (1985) 572.
\bibitem{heat} G.~Heatlie {\em et al.}, Nucl.Phys.{\bf B352}(1991) 266. 
\bibitem{MARTIW}
G.~Martinelli, Phys.~Lett. {\bf B141} (1984) 395;
C.~Bernard {\em et al.},Phys.Rev.{\bf D36} (1987) 3224.
\bibitem{WI}
G.~Martinelli {\em et al.},
Phys.~Lett. {\bf B311} (1993) 241; Erratum {\bf B317} (1993) 660;
M.~Paciello {\em et al.}, Phys.~Lett. {\bf B341} (1994) 187.
\bibitem{JAPBK} JLQCD, S. Aoki {\em et al.}, hep-lat/9705035;
Nucl.~Phys. {\bf B} (Proc. Suppl.) {\bf 53} (1997) 349.
\bibitem{gupta_bp} 
T.~Bhattacharya {\em et al.}, Phys.~Rev. {\bf D55} (1997) 4036.
\bibitem{giusti}C.R.Allton {\em et al.},Nucl.Phys.{\bf B489} (1997) 427.
\end{thebibliography}
\end{document}